\documentclass[11pt]{article}

\usepackage[margin=1in]{geometry}
\usepackage{times}
\usepackage{microtype}
\usepackage{graphicx}
\usepackage{booktabs}
\usepackage{amsmath, amssymb}
\usepackage{siunitx}
\usepackage{enumitem}
\usepackage[hyphens]{url}
\usepackage[numbers,sort&compress]{natbib}
\usepackage[colorlinks=true,allcolors=blue]{hyperref}
\usepackage{tikz}
\usetikzlibrary{shapes.geometric, arrows.meta, positioning, calc}

\title{Clever Hans in Chemistry: Chemist Style Signals Confound Activity Prediction on Public Benchmarks}

\author{Andrew D. Blevins and Ian K. Quigley}
\date{}

\begin{document}
\maketitle

\begin{abstract}
Can machine learning models identify which chemist made a molecule from structure alone? If so, models trained on literature data may exploit chemist intent rather than learning causal structure-activity relationships. We test this by linking \textsc{ChEMBL} assays to publication authors and training a 1,815-class classifier to predict authors from molecular fingerprints, achieving 60\% top-5 accuracy under scaffold-based splitting. We then train an activity model that receives only a protein identifier and an author-probability vector derived from structure, with no direct access to molecular descriptors. This author-only model achieves predictive power comparable to a simple baseline that has access to structure. This reveals a ``Clever Hans'' failure mode: models can predict bioactivity largely by inferring chemist goals and favorite targets without requiring a lab-independent understanding of chemistry. We analyze the sources of this leakage, propose author-disjoint splits, and recommend dataset practices to decouple chemist intent from biological outcomes.
\end{abstract}

\section{Introduction}

In a meeting years ago, an experienced chemist glanced at a panel of compounds in front of us and said ``that's a Stuart Schreiber molecule.'' The observation has haunted us: if humans can recognize chemist style from structure alone, could our machine learning models be doing the same? And if so, are we training models that learn chemistry and biology, or models that simply exploit intent leakage---the stylistic regularities that arise as chemists pursue specific goals?

This failure mode is well documented in computer vision, where it goes by names like ``Clever Hans'' or shortcut learning \citep{lapuschkin2019cleverhans,geirhos2020shortcut}. Models latch onto dataset signatures and spurious correlations that work in aggregate but encode the wrong mechanism \citep{torralba2011unbiased}, resulting in impressive benchmark numbers that collapse once out of distribution. We show that analogous shortcut behavior appears in medicinal chemistry.

\begin{figure}[t]
\centering
\begin{tikzpicture}[
    box/.style={rectangle, draw, rounded corners, minimum width=2.2cm, minimum height=1cm, align=center, font=\small},
    arrow/.style={->, >=stealth, thick},
    stage/.style={font=\small\bfseries, color=blue!70!black}
]
\node[stage] at (-6.5, 4.5) {Typical Activity Model};
\node[box, fill=green!10] (mol0) at (-8, 3.5) {Molecule\\Structure};
\node[box, fill=purple!10] (prot0) at (-5, 3.5) {Protein\\ID};
\node[box, fill=blue!10] (model0) at (-6.5, 2) {Activity\\Model};
\node[box, fill=red!10] (activity0) at (-6.5, 0.5) {Active /\\Inactive};

\draw[arrow] (mol0) -- (model0);
\draw[arrow] (prot0) -- (model0);
\draw[arrow] (model0) -- (activity0);

\node[stage] at (0, 4.5) {Stage 1: Author Identification};
\node[box, fill=green!10] (mol1) at (0, 3.5) {Molecule\\Structure};
\node[box, fill=blue!10] (model1) at (0, 2) {Author ID\\Model\\(GBDT)};
\node[box, fill=orange!10] (auth) at (0, 0.5) {Author\\Probabilities\\(1,815-dim)};

\draw[arrow] (mol1) -- (model1);
\draw[arrow] (model1) -- (auth);

\node[stage] at (6.5, 4.5) {Stage 2: Activity Prediction};
\node[box, fill=orange!10] (auth2) at (5, 3.5) {Author\\Probabilities};
\node[box, fill=purple!10] (prot) at (8, 3.5) {Protein\\ID};
\node[box, fill=blue!10] (model2) at (6.5, 2) {Activity\\Model\\(GBDT)};
\node[box, fill=red!10] (activity) at (6.5, 0.5) {Active /\\Inactive};

\draw[arrow] (auth2) -- (model2);
\draw[arrow] (prot) -- (model2);
\draw[arrow] (model2) -- (activity);

\node[draw, dashed, thick, fill=yellow!10, font=\footnotesize\bfseries, align=center, text width=2.8cm] 
    at (6.5, -0.8) {No direct molecular\\features used!};

\draw[arrow, dashed, color=orange!70] (auth) to[out=0, in=180] (auth2);

\end{tikzpicture}
\caption{Overview of the two-stage chemist style leakage test. \textbf{Stage 1:} We train a model to predict which of 1,815 authors synthesized a molecule from its structure alone, achieving 60\% top-5 accuracy. \textbf{Stage 2:} We use these author probabilities (plus protein ID) to predict activity, \emph{without} providing molecular features. The strong performance (validation AUROC around 0.65) reveals a shortcut: models can predict activity by inferring chemist intent without explicitly modeling structure--activity relationships.}
\label{fig:overview}
\end{figure}
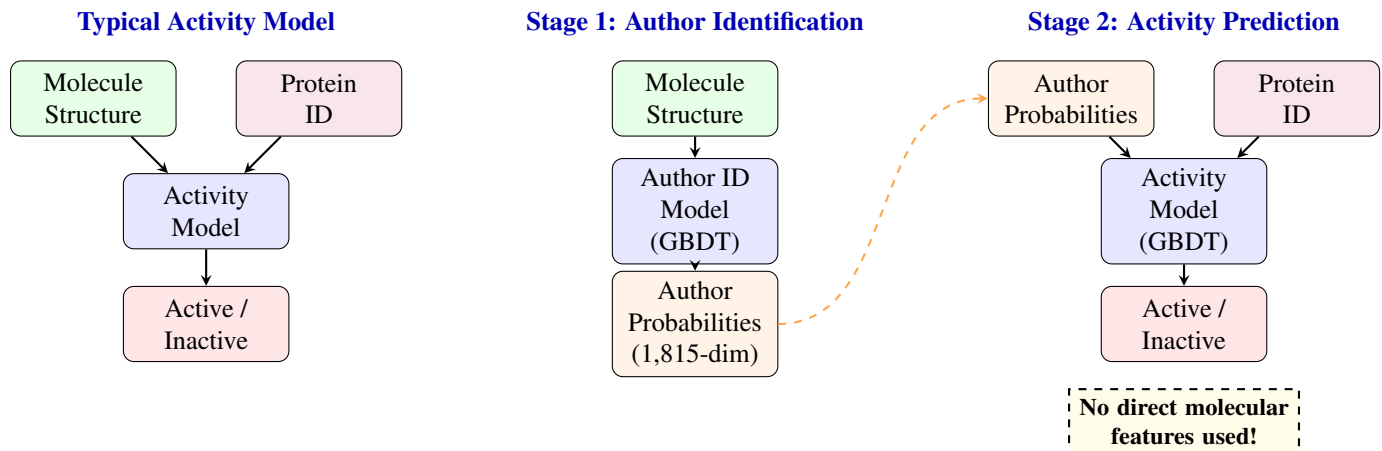

We test this hypothesis through a two-stage experiment (Figure~\ref{fig:overview}). First, we link \textsc{ChEMBL} assays to publication authors, creating an author--molecule graph spanning $1{,}815$ prolific authors. We train a classifier to predict \emph{who} made a molecule from structure alone and find that chemist styles are highly distinctive and readily learnable. Second, we train an activity model that receives only a protein identifier and the predicted author probabilities, with no direct access to molecular structure. This author-only model achieves competitive performance (Table~\ref{tab:activity-results}), comparable to a simple ECFP+protein baseline, showing that activity can be predicted from chemist intent proxies even when direct structural descriptors are withheld.

In this work we use the term \emph{chemist style} to refer to regularities in how molecules are designed within a lab or author group: preferred scaffolds, functional groups, reaction motifs, and target families. Operationally, however, we do not observe individual molecule designers. Instead, we infer ``style'' from \textsc{ChEMBL} document authorship metadata, treating each author identifier as a proxy for a lab- or PI-level group. This authorship-derived notion is imperfect---papers may list many authors, some of whom are biologists or non-synthetic contributors---but it is sufficient to ask whether authorship-linked patterns are strong enough to confound activity prediction.

The problem runs deeper than known curation pitfalls \citep{baell2010pains, fourches2010curation}. Community datasets aggregate heterogeneous literature under uneven experimental intent, and even rigorous standardization efforts \citep{bequignon2023papyrus, sun2017excape} leave residual signals that modern models readily exploit. We show that dataset construction choices encode chemist style signals that confound evaluation, analogous to known shortcuts in computer vision benchmarks, and we point toward leakage-aware splits, adversarial debiasing, and cross-lab replication as mitigations.

\section{Related Work}

A large body of protein-ligand interaction work trains and/or validates on ChEMBL-derived bioactivity or related resources (e.g., BindingDB, KIBA): examples include DeepDTA, GraphDTA, MolTrans, and DeepAffinity \citep{ozturk2018deepdta,nguyen2020graphdta,huang2021moltrans,karimi2019deepaffinity}. Earlier ligand-based target prediction on ChEMBL demonstrated strong performance from classical ML \citep{koutsoukas2013chembltp,mayr2018largescale}, and the ChEMBL resource itself continues to evolve toward higher-quality deposition and tooling \citep{bento2024chembl}. Recent curation efforts (e.g. ExCAPE-DB, Papyrus, MF-PCBA) provide large, cleaned, binarized activity sets meant to reduce assay noise and ease fair benchmarking \citep{papyrus2022, sun2017excape, buterez2023mfpcba}. We situate our work within this lineage but show that author/style signals remain a strong confound.

Structure-first models now predict complex biomolecular assemblies and sometimes claim binding improvements. AlphaFold~3 introduced a diffusion architecture that models proteins, nucleic acids, ligands, and modifications in a unified framework \citep{alphafold3}. Boltz-2 extends this line by modeling both complex structures and (approximate) binding affinities as a joint task \citep{boltz2}. In molecular property prediction, foundation pretraining on massive unlabeled chemical corpora (e.g., Recursion’s MolE) can yield strong ADMET and activity baselines \citep{mole2024}. Our results complement these advances by isolating a distinct failure mode: chemist style shortcuts that may inflate performance irrespective of the underlying biological mechanism.

Extended-Connectivity Fingerprints (ECFPs) \citep{rogers2010ecfp} combined with boosted trees or random forests remain competitive for many ChEMBL-style tasks, particularly under leakage-prone splits \citep{robinson2020reanalyze,riniker2013benchmark2d,hu2012ecfpvs3d,yang2019analyzing}. We use a gradient boosting model on ECFPs as a deliberately simple probe to measure the strength of chemist style signals and to emphasize how non-causal shortcuts can dominate performance in public benchmarks.

\begin{figure}
    \centering
    \includegraphics[width=1.0\linewidth]{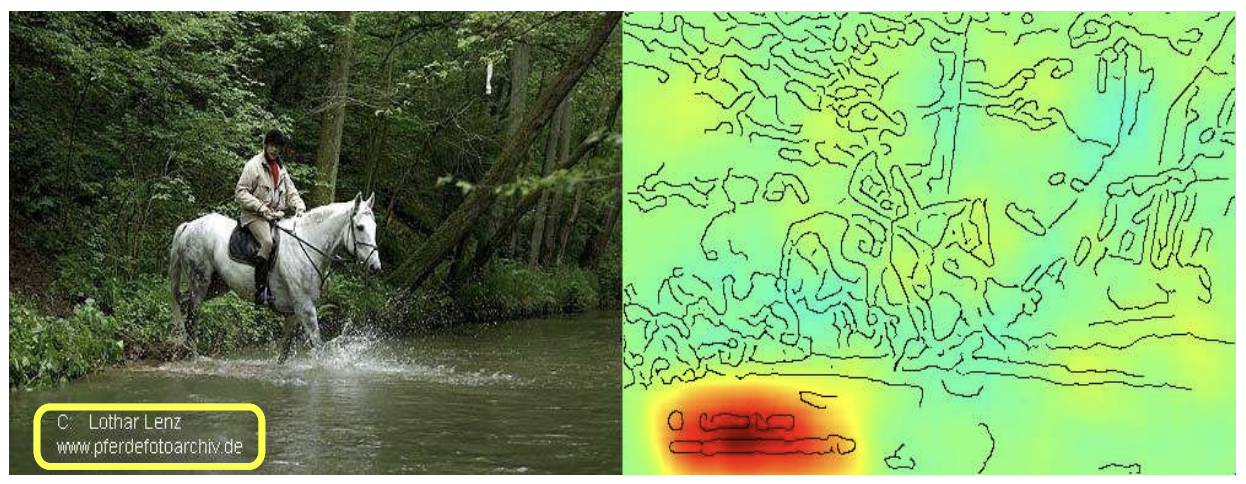}
    \caption{Classic example of a Clever Hans predictor, reproduced from Lapuschkin et al.~\cite{lapuschkin2019cleverhans}. The left panel shows an image correctly classified as containing a horse; the right panel shows a relevance heatmap revealing that the classifier bases its decision primarily on the photographer’s corner watermark rather than on the horse itself. By analogy, in this work we argue that molecules in ChEMBL carry chemist-specific “watermarks” in their structure, which even simple models can exploit, confounding attempts to learn genuine causes of binding}
    \label{fig:cleverhans}
\end{figure}

Computer vision has long documented dataset--specific signals and cross--dataset failures, with a prominent example being \citet{torralba2011unbiased}. Subsequent work formalized \emph{shortcuts}---decision rules that exploit unintended cues and fail under distribution shift \citep{geirhos2020shortcut}---and exposed failure modes via explanation methods \citep{lapuschkin2019cleverhans}. Robustness benchmarks such as ImageNet-C/A \citep{hendrycks2019imagenetc,hendrycks2021imageneta} and the WILDS benchmark for real distribution shifts \citep{koh2021wilds} have driven better evaluations alongside algorithmic responses (e.g., GroupDRO and IRM) \citep{sagawa2019distributionally,arjovsky2019invariant}. Together, these literatures motivate our focus on chemist style leakage as a chemistry--specific shortcut. 

Beyond well-known scaffold leakage and congeneric series effects, several works have documented systematic issues in public chemoinformatics benchmarks, including decoy and analogue bias in structure-based virtual screening (DUD-E) \citep{chen2019dude,wallach2018ave}, target/assay heterogeneity and label mixing \citep{sieg2019biascontrol,landrum2024mixing}, and performance collapses around activity cliffs \citep{vantilborg2022cliffs}. Methodologically, temporal or simulated-temporal splits can better approximate prospective utility than random or naive scaffold splits \citep{sheridan2013timesplit,landrum2023simpd}. We add \emph{authorship} as a concrete, measurable axis of leakage with direct consequences for activity prediction, and we propose author-aware splits and diagnostics as mitigations.

\section{Data}\label{sec:data}

We construct an author--molecule view of \textsc{ChEMBL} by joining assay/compound records to publication metadata and author lists. Each molecule inherits the (multi)set of authors of the linked paper. To obtain well-supported supervision, we restrict to prolific authors (more than 30 publications and more than 600 molecules observed), yielding $1{,}815$ authors presumed to correspond primarily to PI-level chemists or assay leads.

We study two complementary prediction problems:
\begin{enumerate}[leftmargin=1.5em, itemsep=0.25em]
    \item \textbf{Author identification (1{,}815--way).} Input: a small molecule; Output: a probability distribution over authors. Intended to measure the distinctiveness of ``chemist style.''
    \item \textbf{Activity prediction (binary).} Input: a protein identifier and the \emph{author--probability vector} for the molecule produced by Task~1; Output: active vs.\ inactive. No direct molecular descriptors are provided to this model.
\end{enumerate}

\emph{Task~1 (author identification).} We standardize structures, deduplicate, and compute circular fingerprints to represent molecules. Labels are multi-author where applicable; evaluation counts a prediction as correct if the predicted author appears among the paper’s author set (top-$k$ metrics analogously). Implementation specifics (fingerprint parameters, canonicalization, multi-author handling) are in App.~\ref{app:task1-pre}.

\emph{Task~2 (activity prediction).} Starting from \textsc{ChEMBL} bioactivity records linked to the $1{,}815$ authors above, we clean, normalize, and binarize activities via a Papyrus-style protocol, then build examples as compound--target pairs. Features are limited to (i) the Task~1 author-probability vector and (ii) a tokenized protein identifier; no chemical structure is exposed. Curation and featurization details are in App.~\ref{app:task2-pre}.

For Task~1 we perform a \emph{scaffold split} using Bemis--Murcko scaffolds to reduce near-duplicate leakage. For Task~2 we re-join Task~1 predictions to compound--target pairs and apply a split disjoint by scaffold at the compound level. 

\section{Methods}\label{sec:models}

\subsection{Author Identification from Structure}

For the author identification task, we use a standard fingerprint + gradient boosting setup.

Each molecule is standardized (canonical SMILES, salt/solvent removal) and featurized with Morgan/ECFP fingerprints (radius 2, 2048 bits, chirality on). We restrict labels to the $1{,}815$ prolific authors described in Sec.~\ref{sec:data} and treat author prediction as a multiclass problem. A single LightGBM classifier is trained on ECFP features to predict an author distribution over these $1{,}815$ classes, using scaffold-grouped splits for train/validation/test. Full preprocessing and hyperparameters are given in App.~\ref{app:task1-pre} and App.~\ref{app:model-details}.

In addition to the multiclass model, we fit one-vs-rest (OvR) binary probes---one per author---on the same features and splits. These OvR models are used only for analysis (per-author ROC--AUC / average precision distributions) and do not feed into downstream tasks.

For each compound $x$, we retain the multiclass softmax vector
$a(x)\in\mathbb{R}^{1815}$ as an \emph{chemist style prior}. This is the only compound-level representation used by the downstream activity model.

\subsection{Activity Prediction from Author Style}

The activity model operates on chemist style priors and protein identity, without access to molecular structure.

Given a compound $x$ and target protein $t$, we construct an input feature vector by concatenating (i) the author-probability vector $a(x)\in\mathbb{R}^{1815}$ from the Task~1 classifier and (ii) a fixed-dimensional hashed representation of the target identifier. We use a standard feature hasher to map the string target ID to a 1,024-dimensional vector, yielding a $2{,}839$-dimensional input.

Training labels are binary activities derived from \textsc{ChEMBL} using a Papyrus-style curation and binarization pipeline (App.~\ref{app:task2-pre}). We enforce scaffold-disjoint splits at the compound level and train over multiple random GroupShuffleSplit folds, using a held-out validation fold in each run for early stopping. A LightGBM binary classifier is then trained on these concatenated author + target features. All curation details, split construction, and model hyperparameters are described in App.~\ref{app:task2-pre} and App.~\ref{app:model-details}.

The resulting model defines $f_\phi(a(x), t)\in[0,1]$, an estimate of the probability that compound $x$ is active against target $t$ based solely on chemist style priors and protein identity.

\begin{figure}
    \centering
    \includegraphics[width=0.5\linewidth]{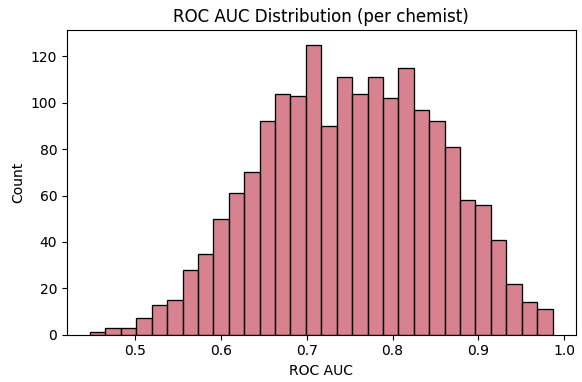}
    \caption{Histogram of ROC AUC scores of each author in one-vs-rest models.}
    \label{fig:ovr_roc}
\end{figure}

\section{Experiments}\label{sec:experiments}

\subsection{Author Identification Results}

\begin{table}[t]
\centering
\caption{Author identification results on the scaffold-split held-out set.}
\label{tab:author-results}
\begin{tabular}{@{}lcccc@{}}
\toprule
\textbf{Model} & \textbf{Top-1} & \textbf{Top-5} & \textbf{Top-10} & \textbf{Log-loss} \\
\midrule
\textbf{ECFP + Gradient Boosting} & \textbf{0.27}  & \textbf{0.60}   & \textbf{0.68}   & \textbf{3.66} \\
Empirical class prior (dummy)     & 0.0003 & 0.0015 & 0.0033 & 7.53 \\
\bottomrule
\end{tabular}
\end{table}

We first ask how well a simple fingerprint--based model can recover who made a molecule from its structure alone. Table~\ref{tab:author-results} shows that the ECFP + GBDT classifier achieves nontrivial top-$k$ accuracies on the 1,815-way author task, whereas a dummy model that only reflects empirical class frequencies is effectively at chance. In other words, even under scaffold-based splitting, chemist style is written clearly enough into molecular structure that a very simple model can reliably distinguish which lab a compound came from.

The one-vs-rest probes in Figure~\ref{fig:ovr_roc} reinforce this picture. Some authors have ROC--AUC scores close to random, but many exhibit highly separable styles: their molecules sit in distinct regions of ECFP space that are easy to carve out with shallow trees. This heterogeneity is exactly what we would expect if different labs pursue characteristic scaffolds, project types, and library designs over long periods of time. From the model’s perspective, the author label is not an incidental annotation; it is a strong, learnable signal embedded in the structures themselves.

\subsection{Author--Only Activity Results}

\begin{figure}
    \centering
    \includegraphics[width=1.0\linewidth]{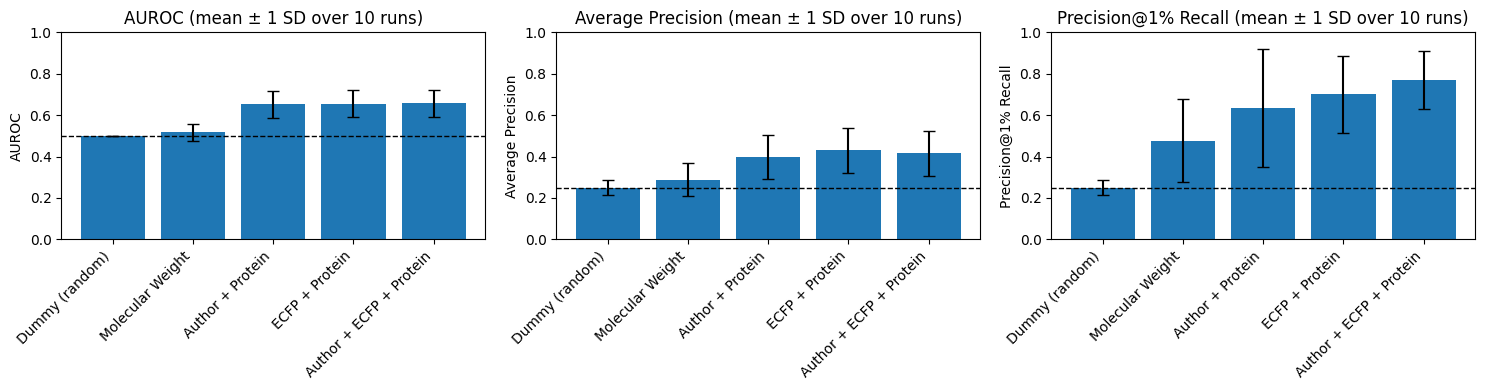}
    \caption{Validation performance of different feature sets under an identical GBDT classifier. Bars show mean performance over five random scaffold splits; error bars denote one standard deviation. The author + protein model tracks the ECFP + protein model close in AUROC, modest gap in AP. Combining author probabilities with ECFPs yields only modest additional gains, indicating that a large fraction of the predictive signal is already present in chemist style and target identity.}
    \label{fig:activity_model}
\end{figure}

We next ask whether these author--style signals are strong enough to support activity prediction without any molecular descriptors. The author--only activity model receives only the author-probability vector $a(x)$ from Task~1 and a hashed protein identifier, but no structural features. Because the activity model only interacts with structure through this fixed author classifier, any structural detail that does not influence predicted author probabilities is effectively invisible at activity training time. In this sense, the author-probability vector is a fixed, author-centric representation of the molecule. It restricts the activity model to whatever structural information is preserved by the author classifier. We compare this to a family of baselines that all use the same LightGBM architecture but differ in which features are exposed (Table~\ref{tab:activity-results}, Figure~\ref{fig:activity_model}).

As expected, a random dummy model and a very simple molecular weight baseline sit close to chance. In contrast, both the author+protein and the ECFP+protein models achieve substantially higher AUROC and precision--recall performance, and---crucially---they are very similar to one another across metrics and across random splits. Adding author probabilities on top of ECFPs yields only modest additional gains. Taken together, this says that a large fraction of the predictive power we usually attribute to ``learning structure--activity relationships'' on this curated \textsc{ChEMBL} slice is also attainable by a model that never sees molecular structure at all, but only needs to understand which authors tend to work on which targets and chemotypes.

\begin{table}[t]
\centering
\caption{Activity prediction results: mean validation AUROC, average precision (AP), and precision at 1\% recall (P@1\%R) over five random scaffold splits. Values are mean $\pm$ standard deviation. Best mean in each column is bold.}
\label{tab:activity-results}
\begin{tabular}{@{}lccc@{}}
\toprule
\textbf{Model / Features} & \textbf{AUROC} & \textbf{AP} & \textbf{P@1\%R} \\
\midrule
Dummy (random) &
$0.500 \pm 0.000$ &
$0.249 \pm 0.036$ &
$0.249 \pm 0.036$ \\
Molecular weight + protein ID &
$0.517 \pm 0.043$ &
$0.288 \pm 0.079$ &
$0.477 \pm 0.202$ \\
Author probs + protein ID &
$0.652 \pm 0.064$ &
$0.399 \pm 0.107$ &
$0.636 \pm 0.285$ \\
ECFP + protein ID &
$0.656 \pm 0.067$ &
$\textbf{0.430} \pm 0.107$ &
$0.700 \pm 0.188$ \\
Author probs + ECFP + protein ID &
$\textbf{0.658} \pm 0.065$ &
$0.415 \pm 0.111$ &
$\textbf{0.769}\pm 0.142$ \\
\bottomrule
\end{tabular}
\end{table}

Put differently, much of the information that a simple structure-based model exploits in this setting is \emph{explainable by chemist style}. The activity model does not need to infer detailed chemistry to perform well; it can instead learn the sociology of the dataset---how different labs behave, which series they pursue, and which targets they favor. This is precisely the Clever Hans failure mode we set out to study: strong benchmark performance that can be driven by intent leakage and lab-specific regularities and may not reflect a causal understanding of structure--activity relationships. As a consequence, we argue that good results on \textsc{ChEMBL}-derived benchmarks should be interpreted with caution unless author-aware splits and baselines are used to rule out this shortcut.

\paragraph{How structural is the author representation?}
As a sanity check on the representation used by the Stage~2 model, we asked how much structural information is encoded in the author-probability vectors themselves. We randomly selected a subset of hashed ECFP bits (512-bit Morgan fingerprints, radius~2) and, for each bit, trained a separate logistic-regression probe to predict the bit from the corresponding author-probability vector on a held-out set of molecules. These simple linear decoders achieve high ROC--AUC values (with median performance around 0.9 across the probed bits) and substantial average-precision lift over the base bit frequencies, indicating that chemist-style predictions retain a rich, highly structured view of chemical space. We interpret this as evidence that public medicinal-chemistry datasets occupy a narrow ``chemist-style'' manifold: once a model has learned to recognize which authors a molecule most resembles, much of its circular-fingerprint representation is already determined. This reinforces our conclusion that apparent structure--activity signal on \textsc{ChEMBL}-derived benchmarks is tightly entangled with chemist style and data provenance.

\section{Conclusion}

We have shown that authorship-linked patterns in \textsc{ChEMBL} act as a strong shortcut for both author identification and activity prediction. A simple ECFP-based classifier can reliably recover which of 1{,}815 prolific authors synthesized a molecule from structure alone, and the resulting author-probability vectors, combined only with protein identifiers, support activity prediction performance that closely tracks an ECFP+protein baseline. Thus, on this benchmark slice, much of the apparent structure--activity signal can be reproduced without exposing molecular descriptors at training time, simply by learning chemist style and target portfolios.

This suggests that source identities---authors, labs, institutions, vendors, or campaigns---are an important and underappreciated confounder in public medicinal chemistry datasets. Whenever labels are drawn from many different sources, models can benefit from predicting \emph{who} produced the data and \emph{what} they typically work on, even if that information was never intended as an input feature.

As a result, we recommend that benchmark designers and practitioners:
(i) retain and report source metadata where possible,
(ii) include source-only or source+target baselines to bound the contribution of source signals, and
(iii) consider source-aware splits (e.g., author-disjoint, lab-disjoint, or site-disjoint) alongside scaffold and temporal splits, particularly for headline numbers. These practices will help distinguish genuine structure--activity learning from performance that is largely driven by shortcuts on the provenance of the data.

\section{Artifacts}

All code, configuration files, and processed data needed to reproduce our experiments are available at
\href{https://github.com/Leash-Labs/clever-hans}{\texttt{github.com/Leash-Labs/clever-hans}}.

We will also release a leaderboard ranking all chemists in our ChEMBL subset by stylistic consistency at:
\url{https://leash-labs.github.io/chemist-style-leaderboard}.

\section*{Acknowledgments}
We thank Michael Cuccarese for inspiring this paper.

We thank the maintainers of \textsc{ChEMBL} and the open--source communities behind LightGBM and RDKit libraries.

\bibliographystyle{plainnat}
\bibliography{refs}
\clearpage
\appendix

\section{Data Cleaning and Preprocessing}\label{app:cleaning}

\subsection{Author Identification (Task~1) Preprocessing}\label{app:task1-pre}

We construct an author--molecule table from \textsc{ChEMBL} v33 using BigQuery. Documents (\texttt{docs\_33}) are joined to activities (\texttt{activities\_33}), then to compound records and structures, yielding (\texttt{doc\_id}, InChI key, SMILES) pairs. 

Author strings are exploded, normalized (lowercased, punctuation-stripped, whitespace-collapsed), and filtered to tokens that look like ``lastname initials.'' These tokens serve as author identifiers. The resulting table is deduplicated per (author, InChI key).

We then compute per-author statistics (distinct molecules and documents) and restrict to \emph{prolific} authors with $n_{\text{molecules}}>600$ and $n_{\text{docs}}>30$, giving 1,815 authors. All Task~1 and Task~2 experiments are performed on molecules associated with this author set.

Molecules are standardized with RDKit (canonical SMILES, salt/solvent removal, parse failures dropped), deduplicated per (author, canonical SMILES), and assigned Bemis--Murcko scaffolds for split grouping. Features are 2048-bit Morgan/ECFP fingerprints (radius 2, chirality on). Labels are multi-author where applicable; a top-$k$ prediction is counted as correct if any true author is in the top-$k$.

Scaffold-grouped splits (using Bemis--Murcko scaffolds as groups) define train/validation/test partitions. The trained multiclass model outputs a softmax vector $a(x)\in\mathbb{R}^{1815}$ for each molecule; this is cached and reused in Task~2.

\subsection{Activity Prediction (Task~2) Preprocessing}\label{app:task2-pre}

For Task~2 we start from \textsc{ChEMBL} activities linked to the 1,815 prolific authors and apply a Papyrus-style curation:

\begin{itemize}[leftmargin=1.5em,itemsep=0.25em]
  \item normalize measurements to $p$-scale (e.g., $p\mathrm{IC}_{50}$, $pK_i$);
  \item keep only exact (non-censored) values;
  \item for each (compound, target, measure-type), aggregate concordant measurements (within 0.5 log units of the pair median);
  \item drop single-assay pairs from assays with poor reproducibility;
  \item binarize with an activity threshold of $p\mathrm{X}\ge 6$.
\end{itemize}

The resulting table includes: InChI key, SMILES, target identifier, binary label, and the precomputed author-probability vector $a(x)$. Targets are represented by a 1024-dimensional feature-hashed encoding of the target ID. For models that use structure, we also compute 1024-bit ECFP fingerprints from SMILES. A simple molecular weight feature (RDKit \texttt{MolWt}) is used for a baseline.

Train/validation splits are scaffold-disjoint. The positive rate is roughly 25\%, with on the order of $3\times 10^5$ training and $8\times 10^4$ validation examples in a typical split.

\section{Model and Training Details}\label{app:model-details}

\subsection{Author Model (Task~1)}

The author identification model is a LightGBM multiclass classifier trained on 2048-bit ECFP features:

\begin{itemize}[leftmargin=1.5em,itemsep=0.25em]
  \item \texttt{objective = multiclass}, \texttt{num\_class = 1815},
  \item \texttt{metric = multi\_logloss},
  \item typical tree parameters: \texttt{num\_leaves = 31}, \texttt{learning\_rate = 0.05}, \texttt{feature\_fraction = 0.8}, \texttt{bagging\_fraction = 0.8}, \texttt{max\_depth = 6}, \texttt{lambda\_l2 = 1.0}, \texttt{is\_unbalance = True}.
\end{itemize}

We train with early stopping on a scaffold-grouped validation fold, up to 2000 boosting rounds. One-vs-rest probes use the same features and splits but with a binary objective and are used only for analysis (Figure~\ref{fig:ovr_roc}).

\subsection{Activity Models (Task~2)}

All non-dummy activity models are LightGBM binary classifiers trained on different feature combinations:

\begin{itemize}[leftmargin=1.5em,itemsep=0.25em]
  \item base parameters: \texttt{objective = binary}, \texttt{n\_estimators = 800}, \texttt{learning\_rate = 0.02}, \texttt{num\_leaves = 63}, \texttt{subsample = 0.8}, \texttt{colsample\_bytree = 0.8};
  \item early stopping with a patience of 50 rounds on the scaffold-disjoint validation split (monitoring AUC).
\end{itemize}

We compare:
\begin{itemize}[leftmargin=1.5em,itemsep=0.25em]
  \item \textbf{Dummy}: random labels;
  \item \textbf{Molecular weight}: uses only the scalar molecular weight feature;
  \item \textbf{Author + protein}: uses $a(x)$ (1815-d) and hashed target ID (1024-d), no structural descriptors;
  \item \textbf{ECFP + protein}: uses 1024-d ECFP and hashed target ID;
  \item \textbf{Author + ECFP + protein}: concatenation of all three blocks.
\end{itemize}

For each model we report AUROC, average precision, and precision at 1\% recall on the validation split (Table~\ref{tab:activity-results}, Figure~\ref{fig:activity_model}). A shuffled-target sanity check (not shown) confirms that performance of the author+protein model depends on genuine structure/author/target relationships rather than trivial label leakage.

\section{Chemist Style Examples}\label{app:chemist-style}

\begin{figure}
    \centering
    \includegraphics[width=1.0\linewidth]{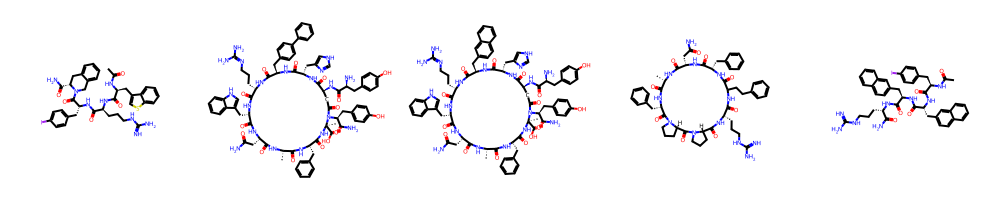}
    \includegraphics[width=0.5\linewidth]{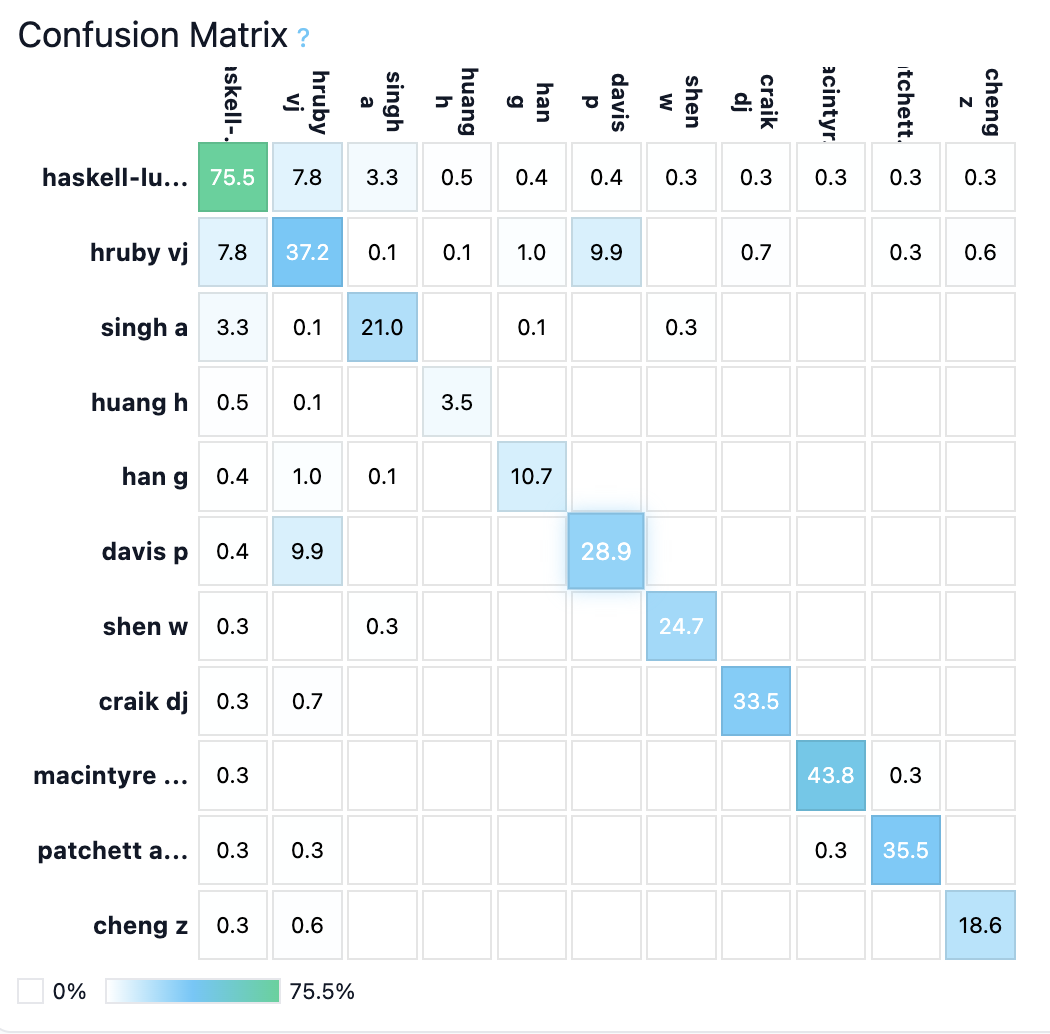}
    \caption{Five exemplar molecules from Carrie Haskell-Luevano and the corresponding author-confusion profile. The top panel shows macrocyclic melanocortin ligands built from densely functionalized, noncanonical amino acids and conformational constraints that are rare elsewhere in \textsc{ChEMBL}. These macrocycle-heavy series make Haskell-Luevano the most stylistically identifiable chemist in our prolific-author cohort: the author classifier assigns her as the top prediction for 75.5\% of held-out molecules under scaffold splitting. The bottom panel shows the confusion-matrix row for this author, with a sharp diagonal peak and only limited spillover onto a small number of chemically similar peptide labs.}
    \label{fig:haskell-molecules}
\end{figure}

\begin{figure}
    \centering
    \includegraphics[width=1.0\linewidth]{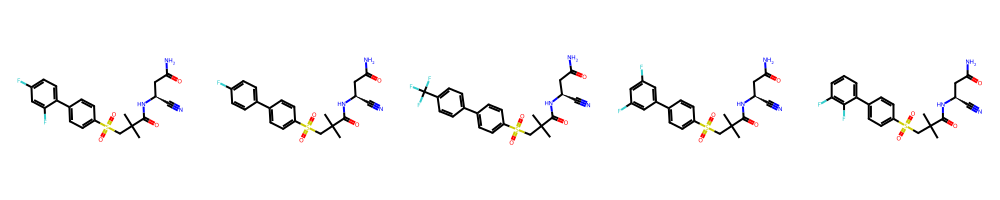}
    \includegraphics[width=0.5\linewidth]{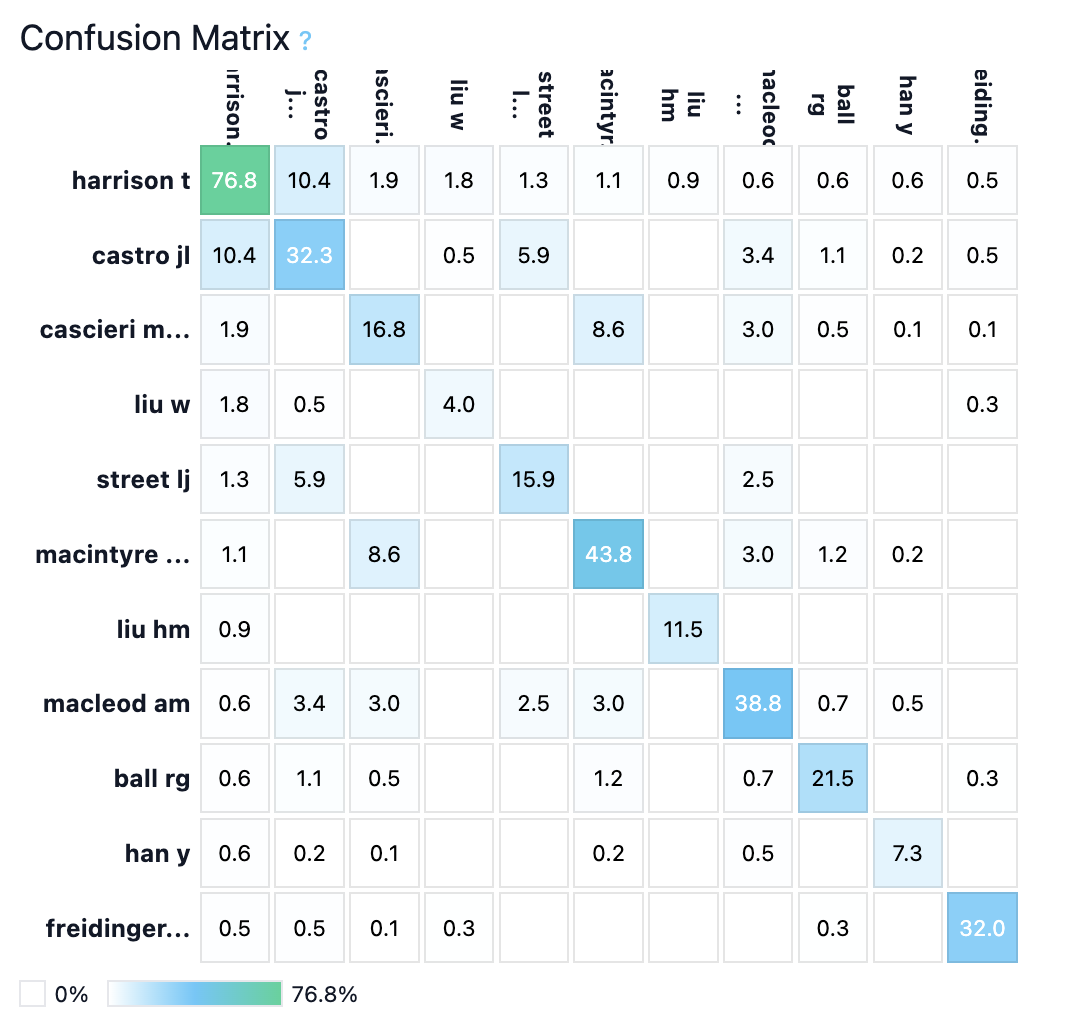}
    \caption{Representative molecules and confusion profile for T.\ Harrison, the third most stylistically identifiable chemist in the prolific-author set. The top panel illustrates tightly focused SAR series built around a shared, highly substituted heterocyclic or spirocyclic core with systematic variation of pendant aryl, heteroaryl, and basic side chains---the sort of NK receptor antagonist campaigns that generate many near-neighbor analogues. This combination of an unusual core scaffold and dense local SAR produces a very distinctive fingerprint signature, so the classifier assigns Harrison correctly on the vast majority of held-out molecules and otherwise confuses him with only a small cluster of nearby authors, as seen in the bottom confusion-matrix row.}
    \label{fig:harrison-molecules}
\end{figure}

\begin{figure}
    \centering
    \includegraphics[width=1.0\linewidth]{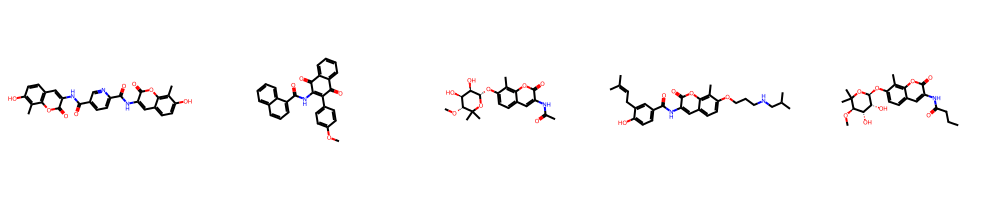}
    \includegraphics[width=0.5\linewidth]{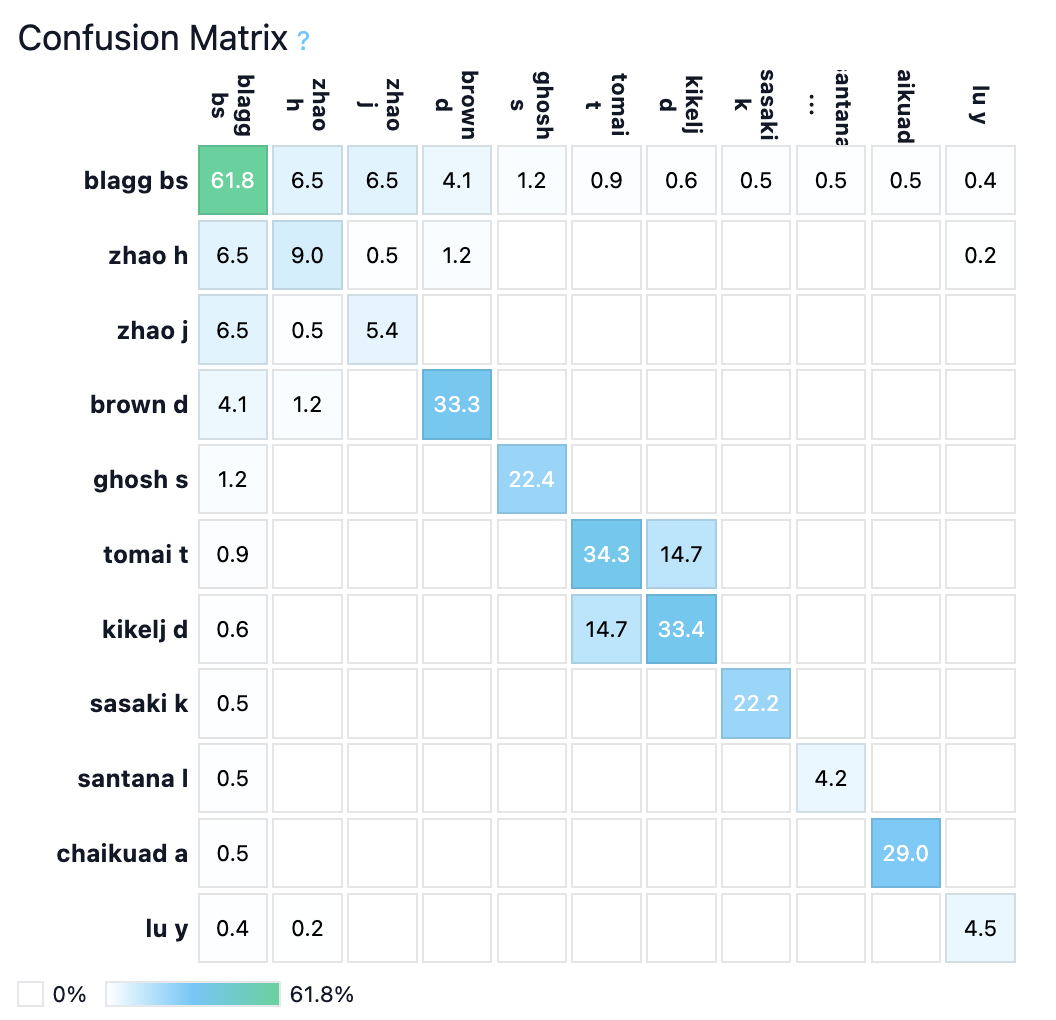}
    \caption{Representative molecules and confusion profile for Brian S.\ J.\ Blagg, ranked sixth by stylistic identifiability in our dataset. The top panel shows novobiocin-inspired Hsp90 C-terminal inhibitors and related analogues: coumarin- or benzamide-like cores coupled to heavily decorated aryl amides and noviose/noviomimetic sugar surrogates. This natural-product-derived chemotype is comparatively uncommon in \textsc{ChEMBL}, and the author classifier therefore predicts Blagg as the source for 68.1\% of his held-out molecules from structure alone. The bottom panel shows the corresponding confusion-matrix row, where errors are concentrated on a small set of other natural-product-like series, reinforcing the idea that the model has learned a highly specific notion of ``Blagg-style'' chemistry.}
    \label{fig:blagg-molecules}
\end{figure}

\begin{figure}
    \centering
    \includegraphics[width=0.5\linewidth]{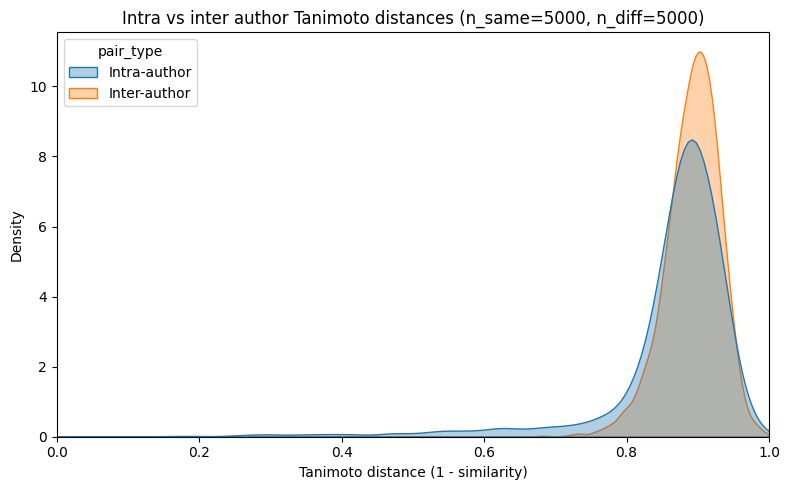}
    \caption{Plotting the Tanimoto distances between molecules that share an author vs those that don't. The ease of prediction of the author dataset cannot be explained by a simple distance metric.}
    \label{fig:inter_intra_dist}
\end{figure}

\end{document}